%% file: main.tex
\documentclass[conference]{IEEEtran}

\input{preamble}      % <- must NOT contain \documentclass

\IEEEoverridecommandlockouts
\input{title_author}

\begin{document}
\maketitle

\input{abstract}
\input{intro}
\input{background}

\input{system_threat}
\input{method}
\input{security}
\input{evaluation}

\input{discussion_conclusion}
\printbibliography
\end{document}

%% file: preamble.tex
\usepackage[style=ieee,sorting=none,maxnames=6,minnames=1,backend=biber]{biblatex}
\usepackage[utf8]{inputenc}
\usepackage[T1]{fontenc}
\usepackage{newtxtext,newtxmath} % Times-like text + math

\usepackage{amsmath,mathtools}
\usepackage{amsthm}

\usepackage{braket}     % \bra, \ket, etc.

\usepackage{siunitx}
\DeclareSIUnit{\mHa}{\milli\hartree}
\DeclareSIUnit{\qubit}{q}

\usepackage{physics}
\makeatletter
\@ifundefined{RenewCommandCopy}{%
  \AtBeginDocument{\let\qty\SI}%
}{%
  \AtBeginDocument{\RenewCommandCopy\qty\SI}%
}
\makeatother

\usepackage[ruled,vlined]{algorithm2e}
\SetKwInput{KwData}{Input}
\SetKwInput{KwResult}{Output}
\DontPrintSemicolon
\SetAlgoNlRelativeSize{-1}
\SetKw{KwRet}{return}

\usepackage{booktabs}   % \toprule, \midrule, \bottomrule
\usepackage{multirow}   % \multirow
\usepackage{adjustbox}  % \begin{adjustbox}{width=\linewidth}
\usepackage{array,tabularx}
\usepackage{graphicx}
\usepackage{xcolor}
\usepackage{float}
\usepackage{placeins}
\usepackage{ragged2e}
\usepackage[inline]{enumitem}
\usepackage{microtype}

\usepackage{tikz}
\usetikzlibrary{arrows.meta,positioning,shapes,backgrounds,fit,calc}

\newtheorem{theorem}{Theorem}[section]
\newtheorem{lemma}{Lemma}[section]

\newtheorem{corollary}{Corollary}[section]

\newcommand{\Dbudget}{\Delta_{\mathrm{budget}}}
\newcommand{\Dkill}{\Delta_{\mathrm{kill}}}
\newcommand{\Dhat}{\hat{\Delta}}                 % use as \Dhat_t
\newcommand{\qrefdist}{q_{\mathrm{ref}}}
\newcommand{\OPIND}{\mathsf{OP\mbox{-}IND}}

\DeclareUnicodeCharacter{00A0}{~}            % NO-BREAK SPACE
\DeclareUnicodeCharacter{2009}{\,}           % THIN SPACE
\DeclareUnicodeCharacter{202F}{\,}           % NARROW NO-BREAK SPACE
\DeclareUnicodeCharacter{207B}{-}            % superscript minus
\DeclareUnicodeCharacter{2076}{^6}           % superscript 6
\DeclareUnicodeCharacter{2264}{\leq}         % ≤
\DeclareUnicodeCharacter{2265}{\geq}         % ≥
\DeclareUnicodeCharacter{2208}{\in}          % ∈
\DeclareUnicodeCharacter{0394}{\Delta}       % Δ
\DeclareUnicodeCharacter{039B}{\Lambda}      % Λ
\DeclareUnicodeCharacter{03BC}{\mu}          % μ
\DeclareUnicodeCharacter{00B5}{\mu}          % µ
\DeclareUnicodeCharacter{2032}{'}            % ′
\DeclareUnicodeCharacter{22A5}{\bot}         % ⟂
\DeclareUnicodeCharacter{2297}{\otimes}      % ⊗

\usepackage[hidelinks,breaklinks=true,bookmarks=false]{hyperref}

%% file: title_author.tex
 \title{Adaptive \textit{t}-Design Dummy-Gate Obfuscation for Cryogenic-Scale Enforcement}

\author{
\IEEEauthorblockN{Samuel Punch}
\IEEEauthorblockA{\textit{School of Computer Science} \\
\textit{University College Cork}\\
Cork, Ireland \\
samuel.punch@ucc.ie}
\and
\IEEEauthorblockN{Krishnendu Guha}
\IEEEauthorblockA{\textit{School of Computer Science} \\
\textit{University College Cork}\\
Cork, Ireland \\
k.guha@ucc.ie}
}

%% file: abstract.tex
\title{Adaptive \textit{t}-Design Dummy-Gate Obfuscation with Auditable Cryogenic Enforcement}

%We give a composable security argument (Abstract-Cryptography style) and \emph{propose—but do not implement—}a zk-SNARK attestation proving that, for each logged interval, the monitor computed $\widehat{\Delta}_t$ and either enforced the policy or aborted, without revealing internal randomness.

\begin{abstract}
Cloud quantum services can reveal circuit structure and timing via scheduler metadata, latency patterns, and co-tenant interference. We introduce \emph{NADGO} (Noise-Adaptive Dummy-Gate Obfuscation), a scheduling \& obfuscation stack that enforces \emph{operational privacy} for gate-model workloads by applying per-interval limits on observable information leakage.
To support confidentiality and fair multi-tenancy, operators need a way to \emph{audit} compliance at acceptable overheads. NADGO combines: (i) hardware-aware $t$-design padding for structured cover traffic; (ii) particle-filter timing randomisation to mask queue patterns; (iii) CASQUE subcircuit routing across heterogeneous backends; and (iv) a per-interval leakage estimator $\widehat{\Delta}_t$ with locked calibration artefacts and a dual-threshold kill-switch. We prototype on a 4-qubit superconducting tile with cryo-CMOS control and evaluate depth-varied local-random circuits and small QAOA instances. Monitoring runs at a $6.3\,\mu$s control interval, and we record per-interval decisions in an append-only, hash-chained audit log.
Across Monte Carlo (Tier~I) and cloud–hardware emulation (Tier~II) evaluations, NADGO keeps $\widehat{\Delta}_t \le \Delta_{\mathrm{budget}}$ in nominal operation (interval-abort SLO $\le 1\%$) and, under attack, yields high AUC separation with concentrated aborts. At matched leakage targets, microbenchmarks indicate lower latency and cryogenic power than static padding, while end-to-end workloads maintain competitive cost envelopes.
\end{abstract}
% No paragraph

%% file: intro.tex
\section{Introduction}
Multi-tenant quantum clouds expose \emph{operational} side channels such as queue timing, scheduler metadata, and co-tenant crosstalk that can reveal circuit structure even when outputs are correct \cite{Fitzsimons2017,Lu2025QuantumLeak,mustafa2024sidechannel-interface}. Such leakage threatens confidentiality and fair multi-tenancy, and persists despite correctness/verification guarantees that focus on outputs rather than operations \cite{Fitzsimons2017}.

\textbf{Challenges.}
Deployable defences must: (i) operate at run time on $\mu$s-scale control loops, (ii) adapt to device drift and heterogeneous back ends, (iii) quantify leakage from operational traces, (iv) support \emph{abort} semantics with auditable logs, and (v) respect latency/power budgets.

\textbf{Prior work and limitations.}
Static padding and fixed-rate batching regularise timing but ignore drift and workload variation; reported mitigations typically act in isolation from compilation, routing, and run-time enforcement \cite{Lu2025QuantumLeak}. Correctness and verifiability protocols protect outputs (and, in blind protocols, algorithm content) rather than the emitted operational metadata \cite{Fitzsimons2017}. State and unitary $t$-designs hide state structure but are rarely integrated with schedulers/routers and lack per-interval enforcement \cite{ambainis2007quantum-tdesigns,Brandao2016LocalDesigns,Haferkamp2022RandomTDesigns}. Noise- and calibration-aware compilation improves fidelity under drift yet does not address timing/metadata side channels \cite{kurniawan2024calibration-aware,Ferrari2022NoiseAdaptiveCompilation}. Interface- and scheduler-level studies further evidence practical leaks in today’s stacks \cite{mustafa2024sidechannel-interface,Lu2025QuantumLeak}.

\textbf{Motivating example.}
A tenant submits circuits of two depths, $D_1<D_2$. An insider observing queue and latency patterns can infer depth via clustering. Our monitor estimates $\widehat{\Delta}_t$ from observable features; if $\widehat{\Delta}_t \ge \Delta_{\mathrm{kill}}$, execution aborts and the event is immutably logged. In \S\ref{sec:evaluation} we show reduced timing correlation versus baselines under the same policy.

\textbf{Our work (NADGO).}
\emph{NADGO} (Noise-Adaptive Dummy-Gate Obfuscation) enforces per-interval leakage limits via: hardware-aware $t$-design padding, drift-adaptive timing randomisation, CASQUE sub-circuit routing, and a locked policy $(\Delta_{\mathrm{budget}},\Delta_{\mathrm{kill}})$ with an automatic kill-switch and audit.

\textbf{Contributions.}
\begin{itemize}
  \item \textbf{Metric \& policy.} An instantaneous estimator $\widehat{\Delta}_t$ for $\Delta I_t$ with a per-interval policy $(\Delta_{\mathrm{budget}},\Delta_{\mathrm{kill}})$ that composes across layers (\S\ref{sec:methods}; Thm.~\ref{thm:opind}).
  \item \textbf{Orchestration.} The NADGO{+}CASQUE pipeline combining $t$-design padding, particle-filter timing randomisation, CASQUE sub-circuit routing, and a monitored kill-switch; the per-job procedure is given in Algorithm~\ref{alg:nadgo}.
  \item \textbf{Evaluation.} A 4-qubit cryo-CMOS prototype with a $6.3\,\mu$s control interval; we report end-to-end overheads and leakage control versus baselines under identical policies (Section~\ref{sec:evaluation}).
\end{itemize}

This article is organised as follows. Section~\ref{sec:background} reviews operational leakage and related work.
Section~\ref{sec:system-threat} states the system and threat model.
Section~\ref{sec:methods} details the estimator, padding, timing, routing, and orchestration.
Section~\ref{sec:evaluation} reports empirical results, with Section~\ref{sec:conclusion} discussing limitations and outlook.

%% file: background.tex
\section{Background and Related Work}
\label{sec:background}

\textbf{Operational leakage and side-channels.}
Multi-tenant control planes leak via timing traces, scheduler metadata, and co-tenant interference—signals that persist even when outputs are correct~\cite{Fitzsimons2017}. On superconducting stacks, bias-current fluctuations in SFQ--DC converters and SQUID arrays yield distinctive profiles that reveal control events~\cite{mustafa2024sidechannel-interface}. At the orchestration layer, Lu~\emph{et al.} show current clouds expose circuit structure through pulse cadence, calibration events, and job descriptors; proposed fixes (e.g., randomised scheduling) act in isolation from compilation, routing, and run-time enforcement~\cite{Lu2025QuantumLeak}. These realities motivate \textsc{nadgo}’s unified defence—$t$-design padding, drift-adaptive timing, topology-aware routing, and \emph{per-interval} policy enforcement; a structured comparison appears in Table~\ref{tab:related-nadgo}.

\textbf{Leakage metric and run-time policies.}
We use an instantaneous leakage metric $\Delta I_t$ with a per-interval policy $(\Delta_{\mathrm{budget}}, \Delta_{\mathrm{kill}})$ to prevent burst leakage being hidden by averages.
Following Mardia~\emph{et al.}~\cite{Mardia2019ConcentrationInequalities}, sharp concentration bounds for empirical discrete distributions improve reliability over short windows, enabling low false-positive kill-switch triggers while retaining burst sensitivity.

\textbf{$t$-design padding and timing randomisation.}
State $t$-designs—$\varepsilon$-approximate ensembles matching the first $t$ Haar moments—provably hide state structure from polynomial-time adversaries~\cite{ambainis2007quantum-tdesigns}.
Ambainis--Emerson show $t{\ge}4$ designs are indistinguishable from Haar-random states; Haferkamp~\emph{et al.}~\cite{Haferkamp2022RandomTDesigns} give tighter depth bounds, and Brandão--Harrow--Horodecki~\cite{Brandao2016LocalDesigns} extend to local random circuits.
\textsc{nadgo} uses hardware-aware, $t{\ge}4$ padding with native-gate constraints to fit coherence budgets, integrating timing perturbations to obfuscate execution cadence, which was not the focus of prior works.

\textbf{Calibration-aware compilation and drift mitigation.}
Noise non-stationarity limits “latest-only” calibration snapshots.
Kurniawan~\emph{et al.}~\cite{kurniawan2024calibration-aware} show historical windows improve fidelity; Ferrari--Amoretti~\cite{Ferrari2022NoiseAdaptiveCompilation} adapt compilation to heavy-hex layouts.
\textsc{nadgo} extends these noise-aware features to operational privacy, feeding calibration trends into a drift-aware scheduler for topology- and timing-randomised routing.

\textbf{Noise-adaptive co-search.}
QuantumNAS~\cite{wang2022quantumnas} couples circuit topology search with qubit mapping for noise robustness. In \textsc{nadgo}, similar noise-awareness also informs dummy padding and timing randomisation for joint fidelity/privacy optimisation.

\textbf{Cryogenic hardware characterisation and control.}
Cryo-CMOS control electronics at mK--few-K temperatures require accurate modelling.
Pérez-Bailón~\emph{et al.}~\cite{PerezBailon2023CryoCMOSMeasurement} and Eastoe~\emph{et al.}~\cite{Eastoe2025LargeScaleCryoCMOS} provide measurement and modelling techniques informing \textsc{nadgo}'s latency budgets and leakage/noise models for sub-10~$\mu$s enforcement.

\textbf{Phase-based obfuscation for compiler threats.}
OPAQUE~\cite{Rehman2025Opaque} hides keys in $R_Z$ rotation angles for IP protection, but does not address timing or metadata leakage; it is complementary to \textsc{nadgo}.

\textbf{Scope and comparison.}
Table~\ref{tab:related-nadgo} summarises representative approaches against the four challenges we target (run-time overhead management, drift adaptation, topology awareness, and operational privacy), and situates \textsc{nadgo} alongside prior work.

\textbf{Positioning.}
\textsc{nadgo} targets operational privacy for gate-model quantum clouds, integrating $t$-design padding, drift-adaptive timing, topology-aware routing, and auditable kill-switch monitoring.
Unlike prior art, which addresses only subsets of Table~\ref{tab:related-nadgo}, the proposed framework of \textsc{nadgo} addresses and unifies all four under dynamic, multi-tenant conditions.

% --- slimmer table, larger font, abbreviated to avoid overflow ---
\providecommand{\cmark}{Y}
\providecommand{\xmark}{--}

\begingroup
\small
\setlength{\tabcolsep}{3.5pt}   % slightly tighter columns
\renewcommand{\arraystretch}{1.05}

\begin{table}[!t]
\centering
\caption{Representative approaches vs.\ key challenges (\cmark{} = addressed; \xmark{} = not in scope).}
\label{tab:related-nadgo}

\begin{tabular}{@{}lccccc@{}}
\hline
\textbf{Framework}
& \textbf{\shortstack{RT\\Overhd}}
& \textbf{\shortstack{Drift\\Adapt.}}
& \textbf{\shortstack{Topo.\\Aware}}
& \textbf{\shortstack{OP\\Priv.}}
& \textbf{Scale} \\
\hline
Scheduler leakage~\cite{Lu2025QuantumLeak}                      & \xmark & \xmark & \xmark & \cmark & cloud \\
State $t$-designs (A\&E'07)~\cite{ambainis2007quantum-tdesigns} & \xmark & \xmark & \xmark & \cmark & theory \\
Calib.-aware~\cite{kurniawan2024calibration-aware}              & \xmark & \cmark & \xmark & \xmark & 7--127 q \\
Noise-adapt. comp.~\cite{Ferrari2022NoiseAdaptiveCompilation}   & \xmark & \cmark & \cmark & \xmark & SC (bench) \\
Cryo-CMOS (scale)~\cite{Eastoe2025LargeScaleCryoCMOS}           & \cmark & \xmark & \cmark & \xmark & large \\
OPAQUE~\cite{Rehman2025Opaque}                                  & \xmark & \xmark & \xmark & \cmark & 20--50 q \\
\hline
\textbf{This work: \textsc{nadgo}}                              & \cmark & \cmark & \cmark & \cmark & 4--32 q (proto) \\
\hline
\end{tabular}

\raggedright
\vspace{2pt}
\emph{Run-time overhead}: actively manages latency/effort;\;
\emph{Drift adaptivity}: adapts to time-varying noise;\;
\emph{Topology awareness}: placement/routing respects hardware;\;
\emph{Operational privacy}: timing/metadata protections.
\end{table}
\endgroup

%% file: system_threat.tex
\section{System and Threat Model}
\label{sec:system-threat}
\begin{figure}[t]
  \centering
  \resizebox{0.92\linewidth}{!}{%  <<< was \linewidth
  \begin{tikzpicture}[
    font=\small,
    every node/.style={outer sep=0pt},
    box/.style={
      draw, rounded corners=1mm, align=center,
      text width=32mm, minimum height=6mm, inner sep=2pt, fill=blue!10
    },
    side/.style={
      draw, rounded corners=1mm, align=center,
      text width=28mm, minimum height=8mm, inner sep=2pt, fill=blue!5
    },
    outbox/.style={
      draw, rounded corners=1mm, align=center,
      text width=32mm, minimum height=8mm, inner sep=2pt, fill=green!15
    },
    flow/.style={-{Stealth[length=3.5pt]}, semithick, gray!70},
    fb/.style={-{Stealth[length=3.5pt]}, dashed, red!70, semithick,
               shorten >=2pt, shorten <=2pt},
    elab/.style={font=\scriptsize, inner sep=0.6pt, fill=white,
                 rounded corners=0.5pt},
    node distance=5mm
  ]
    % Main vertical chain
    \node[box] (client) {Client $C$ \& policy};
    \node[box, below=of client] (tpad) {\textsc{nadgo} $t$-design padder};
    \node[box, below=of tpad] (sched) {Particle-filter scheduler};
    \node[box, below=of sched] (router) {CASQUE router};
    \node[box, below=of router] (mon) {$\hat{\Delta}_t$ monitor \& kill-switch};
    \node[outbox, below=of mon] (out) {Results (policy-compliant)};

    \draw[flow] (client) -- (tpad);
    \draw[flow] (tpad) -- (sched);
    \draw[flow] (sched) -- (router);
    \draw[flow] (router) -- (mon);
    \draw[flow] (mon) -- (out);

    % Side components
    \node[side, right=7mm of router] (hw) {Back-end\\(QPU / TN / CPU / GPU)};
    \node[side, right=7mm of mon] (audit) {Immutable\\audit log};

    \draw[flow] (router) -- node[elab,above]{execute} (hw);
    \draw[flow] (hw) -- node[elab,above]{shots/results} (mon);
\draw[flow] (mon) -- node[elab,above]{events} (audit);

    % Feedback
    \draw[fb] (mon.west) -- ++(-1mm,0)
      |- node[pos=0.35,left,align=left,font=\scriptsize]{leakage\\feedback}
      (tpad.west);
  \end{tikzpicture}%
  }
  \caption{Vertical system pipeline: policy-aligned compilation, secure padding,
           adaptive scheduling/routing, and monitored execution with
           leakage-bound enforcement and audit logging.}
  \label{fig:system-overview}
\end{figure}
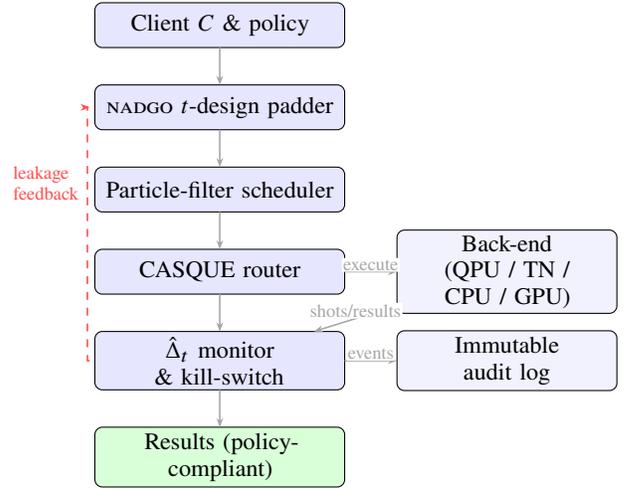

\begin{figure}[t]
  \centering
  \resizebox{\linewidth}{!}{%
  \begin{tikzpicture}[
    font=\small,
    >=Stealth,
    node distance=5mm,
    every node/.style={align=center, outer sep=0pt},
    adv/.style={draw, rounded corners=1mm, fill=red!10,
                text width=34mm, minimum height=8mm, inner sep=1.6pt},
    surfE/.style={draw, rounded corners=1mm, fill=yellow!18,
                  text width=34mm, minimum height=8mm, inner sep=1.6pt},
    mit/.style={draw, rounded corners=1mm, fill=green!14,
                text width=34mm, minimum height=8mm, inner sep=1.6pt},
    arrowT/.style={-Stealth, semithick, gray!65, shorten >=1pt, shorten <=1pt},
    arrowM/.style={-Stealth, semithick, black!75, shorten >=1pt, shorten <=1pt},
    tag/.style  ={draw, fill=gray!15, rounded corners=0.6pt,
                  inner sep=0.6pt, font=\scriptsize}
  ]
    \node[font=\bfseries\small, text width=72mm, minimum height=7mm, align=center] (ttl)
      {Tier~I scope ($n\!\in\!\{4,8,16\}$): evaluated A$\!\to\!$T$\!\to\!$M mappings};
    
    % Midline below title to place two parallel columns
\coordinate (mid) at ($(ttl.south)+(0,-6mm)$); % 5mm gap

    % === Left column: Chain 1 ===
    \node[adv]   (A1) at ($(mid)+(-24mm,0)$) {A1: Scheduler insider\\(timing bursts)};
    \node[surfE, below=of A1] (T1) {T1: Queue timing / admission};
    \node[tag, anchor=north east] at ([xshift=-0.8mm,yshift=1mm]T1.north east) {Control};
    \node[mit,   below=of T1] (M1) {M1: Timing randomisation \& rate limits};
    \draw[arrowT] (A1) -- (T1);
    \draw[arrowM] (T1) -- (M1);

    % === Right column: Chain 2 ===
    \node[adv]   (A2) at ($(mid)+(24mm,0)$) {A2: Co-tenant on QPU\\(RL batch-size modulation)};
    \node[surfE, below=of A2] (T2) {T2: Reschedule / crosstalk hints};
    \node[tag, anchor=north east] at ([xshift=-0.8mm,yshift=1mm]T2.north east) {Execution};
    \node[mit,   below=of T2] (M2) {M2: \textsc{nadgo} dummy gates \& layout jitter};
    \draw[arrowT] (A2) -- (T2);
    \draw[arrowM] (T2) -- (M2);
  \end{tikzpicture}%
  }
  \caption{Threat model (evaluated scope): two parallel chains, A1$\to$T1$\to$M1 (Control) and A2$\to$T2$\to$M2 (Execution).}
  \label{fig:threat-model}
\end{figure}
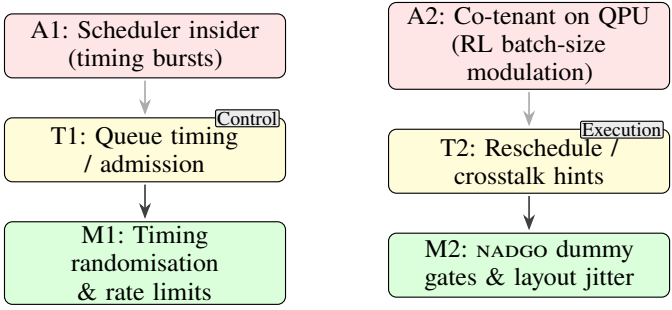

We enforce a per-interval leakage policy \((\Delta_{\mathrm{budget}},\Delta_{\mathrm{kill}})\) with a quantum--classical pipeline that adapts to hardware drift and aborts if estimated leakage exceeds policy limits. The pipeline has four stages—policy-aligned compilation, secure padding, adaptive scheduling/routing, and monitored execution with audit—with feedback to retune padding on-line (Fig.~\ref{fig:system-overview}). The threat model (Fig.~\ref{fig:threat-model}) maps adversaries and attack surfaces to the mitigations evaluated in our experiments.

\subsection*{System Overview}
Client-side compilation (\emph{Qiskit}/\emph{tket}) produces policy-aligned circuits. The \textsc{NADGO} $t$-design padder injects dummy gates to provide cover traffic. A particle-filter scheduler adapts dispatch under drift; CASQUE routing spreads segments across a heterogeneous back-end. The \(\hat{\Delta}_t\) monitor enforces thresholds in real time, aborting on violations and appending to a write-once audit log. Leakage feedback tunes padding parameters.

% Make smaller

\subsection*{Threat Model}
Our goal is to maintain \(\hat{\Delta}_t \le \Delta_{\mathrm{budget}}\) during nominal operation, abort when \(\hat{\Delta}_t \ge \Delta_{\mathrm{kill}}\), and record verifiable audit events. Adversaries can observe control-plane timing, scheduling behaviour, and coarse telemetry, but cannot compromise the policy engine or perform invasive hardware attacks.

\paragraph*{Adversaries and evaluated mitigations.}
We consider two evaluated classes (A1–A2) mapped to attack surfaces (T1–T2) and mitigations (M1–M2). Tier~I evaluates A1$\!\to$T1 and A2$\!\to$T2; telemetry-side channels are out of experimental scope.

% Preamble: \usetikzlibrary{positioning,arrows.meta,calc}

%% file: method.tex
\section{Methodology}
\label{sec:methods}

We formalise the leakage-control objective, specify each \textsc{NADGO} layer (padding, timing, routing, monitoring), and give an operational-privacy (OP-IND) bound. Calibration choices are fixed and auditable to enable reproducibility.

\subsection{Leakage Objective and Estimator}
Let $U_{\mathrm{real}}$ be the hidden computation, $U_{\mathrm{obs}}$ observable traces (timing/metadata/telemetry), $\Lambda_C$ the crosstalk channel, and $Y$ public side information. We measure instantaneous leakage as
\begin{equation}
\Delta I_t \triangleq \sup_{A \in \mathcal{A}_{\mathrm{poly}}} I\!\big(U_{\mathrm{real}}; A(U_{\mathrm{obs}})\,\big|\, \Lambda_C,Y\big),
\label{eq:deltaIt}
\end{equation}
and enforce a dual-threshold policy $(\Delta_{\mathrm{budget}},\Delta_{\mathrm{kill}})$: continue if $\Delta I_t \!\le\! \Delta_{\mathrm{budget}}$, abort if $\Delta I_t\!\ge\!\Delta_{\mathrm{kill}}$.

\paragraph{Estimator.}
We estimate $\Delta I_t$ via a divergence to a locked $t$-design reference with a crosstalk penalty:
\begin{equation}
\widehat{\Delta}_t \;=\; \min\!\left\{ D_{\mathrm{KL}}\!\big(\widehat{p}_t \parallel p^{(\mathrm{des})}_t\big) + \beta_{\mathrm{est}}\|\Lambda_C\|_F,\; \Delta_{\mathrm{kill}} \right\},
\label{eq:estimator}
\end{equation}
where $p^{(\mathrm{des})}_t$ is the design-only reference (per job size and segment class), $\widehat{p}_t$ is the empirical cadence distribution, and $\beta_{\mathrm{est}}$ is a fixed weight.

\paragraph{Cadence features and assumptions.}
We extract a compact feature alphabet $\mathcal{S}$ each interval: inter-dispatch gap $\Delta t$, batch-size ratio $b$ (normalised to a design baseline), queue state $q$ (categorical), and a scalar telemetry proxy $\zeta$. We assume (A1) feature sufficiency for $U_{\mathrm{obs}}$; (A2) Lipschitz crosstalk impact; (A3) conditional independence across intervals given $(\Lambda_C,Y)$.

\paragraph{Calibration (windowing, quantisation, smoothing; locked across tiers).}
To make \eqref{eq:estimator} reproducible, we fix the following pipeline:
\begin{itemize}[leftmargin=1em,topsep=0.2em,itemsep=0.2em]
\item \emph{Windowing:} sliding-window length $W\!=\!128$ intervals; stride $S\!=\!64$ (half overlap). We form a window histogram over codewords and apply an EMA with half-life $H\!=\!10$ windows.
\item \emph{Normalisation:} $\widetilde{\Delta t}\!=\!\Delta t/\mu_{\Delta t}(n)$, $\widetilde{b}\!=\!b/\mu_b(n)$ using design-only baselines for job size $n$.
\item \emph{Vector quantisation (fixed codebook):} bins per feature
\begin{align*}
B_{\Delta t} &= 16~(\text{design quantiles}),\\
B_b          &= 5~\bigl\{\le 0.5,\,(0.5,0.9],\,(0.9,1.1],\,(1.1,1.5],\,>1.5\bigr\},\\
B_q          &= 4~(\textsf{idle},\,\textsf{light},\,\textsf{moderate},\,\textsf{heavy}),\\
B_{\zeta}    &= 8~(\text{design quantiles}).
\end{align*}
This yields $B=16{\times}5{\times}4{\times}8=2560$ codewords. The mapping from $(\widetilde{\Delta t},\widetilde{b},q,\zeta)$ to an index $c_u\!\in\!\{1,\ldots,B\}$ is fixed at calibration.
\item \emph{Smoothing:} Jeffreys add-$\alpha$ with $\alpha=\tfrac{1}{2}$ for both reference and empirical histograms; an EMA combines successive windows. We also mix a tiny uniform mass $\lambda=10^{-3}$ into the design reference to guarantee strict positivity.
\item \emph{Locked reference:} for each $(n,\text{segment})$ class we collect $M\!\approx\!10^6$ design-only intervals to build $p^{(\mathrm{des})}_{n,k}$; we persist $\{p^{(\mathrm{des})}_{n,k}\}$, bin edges, and $(W,S,H,\alpha,\lambda)$ with a checksum. Online, we select $p^{(\mathrm{des})}_t$ by $(n,k)$ and compute $\widehat{p}_t$ with the same codebook.
\end{itemize}

\begin{lemma}[Estimator upper bound]
\label{lem:klbound}
Under (A1)--(A3) and mild regularity of $\Lambda_C$, 
\[
\sup_{A \in \mathcal{A}_{\mathrm{poly}}} I\!\big(U_{\mathrm{real}};A(S_t)\,\big|\,\Lambda_C,Y\big) \;\le\; D_{\mathrm{KL}}\!\big(\widehat p_t\parallel p^{(\mathrm{des})}_t\big) + \beta_{\mathrm{est}}\|\Lambda_C\|_F.
\]
\end{lemma}

\subsection{Noise-Adaptive \texorpdfstring{$t$}{t}-Design Padding}
We insert $\varepsilon_{\mathrm{des}}$-approximate $t$-design layers ($t=\lceil \log_2 n\rceil$, $\varepsilon_{\mathrm{des}}=0.02$) using native Clifford scaffolds. Depth is $\mathcal{O}(n\log n)$ with hardware-aware templates to minimise compilation overhead. The padding both randomises observable cadence and aligns window statistics with $p^{(\mathrm{des})}$.

\subsection{Particle-Filter Timing Randomisation}
Padding alone cannot hide queue cadence. We therefore jitter dispatch times $\theta_k$ using a queue-aware particle filter:
\[
\theta_k \sim \mathrm{PF}(\theta_{k-1}; \sigma_t,\ell_{\max},q_k),\qquad 
\eta^{(i)}\!\sim\!\mathcal{N}\!\big(0,\sigma^2_{\mathrm{proc}}(q_k)\big),
\]
with proposals $\theta^{(i)}_{k}\!\leftarrow\!\mathrm{clamp}(\theta^{(i)}_{k-1}+\eta^{(i)},\pm\sigma_t)$ and weights set to $0/1$ by an $\ell_{\max}$ latency constraint. We trigger resampling when the effective sample size $\mathrm{ESS}\!<\!N/2$. The jitter variance $\sigma_{\mathrm{proc}}(\cdot)$ widens with backlog~$q_k$.

\begin{algorithm}[t]
\caption{PF-TimingSampler (queue-aware jitter control)}
\label{alg:pf}
\DontPrintSemicolon
\KwIn{Particles $\{\theta^{(i)}_{k-1},w^{(i)}_{k-1}\}_{i=1}^N$; base jitter $\sigma_t$; max-latency $\ell_{\max}$; queue state $q_k$}
\KwOut{Updated particles $\{\theta^{(i)}_{k},w^{(i)}_{k}\}_{i=1}^N$}
\For{$i\!=\!1$ \KwTo $N$}{
  $\eta^{(i)} \sim \mathcal{N}\!\big(0,\sigma^2_{\mathrm{proc}}(q_k)\big)$;\;
  $\theta^{(i)}_{k} \leftarrow \mathrm{clamp}(\theta^{(i)}_{k-1}+\eta^{(i)},\pm \sigma_t)$;\;
  $w^{(i)}_{k} \leftarrow w^{(i)}_{k-1}\cdot \mathbf{1}\big[\textsc{Latency}(\theta^{(i)}_{k},q_k)\le \ell_{\max}\big]$;\;
}
Normalise weights;\; \If{$\mathrm{ESS} < N/2$}{Resample and set $w^{(i)}_{k}\!\leftarrow\!1/N$}
\end{algorithm}

\subsection{CASQUE Sub-Circuit Routing}
Given padded segments $\{C'_k\}$, we select a backend $h$ by minimising a multi-objective cost
\begin{equation}
\begin{aligned}
\mathrm{Cost}(h)
= &\; \alpha\bigl(1-\widehat{\mathrm{Fid}}(h)\bigr)
  + \omega_{\mathrm{leak}}\,\mathrm{LeakRisk}(h) \\
  &\; + \gamma\,\mathrm{QueuePenalty}(h).
\end{aligned}
\end{equation}
where $\widehat{\mathrm{Fid}}$ is a hardware-calibrated fidelity proxy, $\mathrm{LeakRisk}$ uses $\widehat{\Delta}_t$ as a signal, and $\mathrm{QueuePenalty}$ scores tail pressure. We switch if $\widehat{\Delta}_t \ge \Delta_{\mathrm{budget}}-\tau$ for $m$ consecutive intervals or when a candidate yields a relative cost drop $\ge\delta$. (Weights $\alpha,\omega_{\mathrm{leak}},\gamma$ and hysteresis $(\tau,m,\delta)$ are fixed across experiments.)

\subsection{Monitoring, Kill-Switch, and Audit}
A lightweight monitor updates $\widehat{\Delta}_t$ each interval and issues an abort if $\widehat{\Delta}_t\!\ge\!\Delta_{\mathrm{kill}}$, completing in $<\SI{5}{\micro\second}$. All decisions (segment, backend, jitter, thresholds, $\widehat{\Delta}_t$) are hashed into an append-only audit log. 

\subsection{Orchestration (Per-Job)}
Algorithm~\ref{alg:nadgo} summarises the per-job control flow and its calls to PF timing (Alg.~\ref{alg:pf}).
\begin{algorithm}[t]
\caption{\textsc{NADGO} Orchestration (per job)}
\label{alg:nadgo}
\DontPrintSemicolon
\KwIn{Circuit $C$; locked artefacts $(q_{\mathrm{ref}},p^{(\mathrm{des})},\text{codebook})$; thresholds $(\Delta_{\mathrm{budget}},\Delta_{\mathrm{kill}})$}
\KwOut{Results or abort; append-only audit log}
$C' \leftarrow \textsc{InsertTDesign}(C)$;\quad Segment $C' \to \{C'_k\}_{k=1}^K$;\;
\For{$k \!=\! 1$ \KwTo $K$}{
  $\theta_k \leftarrow \textsc{PF\_Propose}(q_k)$ \tcp*{Alg.~\ref{alg:pf}}
  $h_k \leftarrow \textsc{SelectBackend}(C'_k)$ \tcp*{CASQUE cost minimisation}
  \textsc{Execute}$(C'_k, h_k, \theta_k)$;\quad $\widehat{\Delta}_t \leftarrow \textsc{UpdateEstimator}()$;\;
  \textsc{Log}$\langle k,h_k,\theta_k,\widehat{\Delta}_t\rangle$;\;
  \If{$\widehat{\Delta}_t \ge \Delta_{\mathrm{kill}}$}{\textsc{AbortAndAttest}(); \Return}
}
\textsc{FuseResults}(); \textsc{FinalizeLog}(); \Return
\end{algorithm}

\subsection{Operational-Privacy Game and Bound}
\paragraph{OP-IND (pointer).}
We adopt the OP-IND game defined above; the privacy bound is stated in Theorem~\ref{thm:opind} (Sec.~\ref{sec:security}).

\paragraph{Notes on reproducibility.}
All estimator hyperparameters $(W,S,H,\alpha,\lambda,B_{\cdot})$, codebook edges, and $\{p^{(\mathrm{des})}_{n,k}\}$ are persisted with a checksum and reused unchanged across tiers. Routing/monitoring hyperparameters $(\alpha,\omega_{\mathrm{leak}},\gamma,\tau,m,\delta)$ are fixed once and remain unchanged across experiments.

%% file: security.tex
\section{Security Argument}
\label{sec:security}

We assess \emph{operational privacy} via the $\OPIND$ game: the adversary chooses $C_0,C_1$ with $\mathsf{pub}(C_0)=\mathsf{pub}(C_1)$; orchestration (Sec.~\ref{sec:system-threat}) runs under $(\Dbudget,\Dkill)$ and reveals the transcript $\tau$ (timing/metadata and abort flag). Let $\Dhat_t$ denote the estimator from Sec.~\ref{sec:methods}, and let $\mathcal{T}$ be the set of admitted (non-aborting) intervals.

\begin{theorem}[OP-IND advantage]
\label{thm:opind}
Assume $\Dhat_t\le\Dbudget(t)$ for all $t\in\mathcal{T}$, calibration $|\mathbb{E}[\Dhat_t]-\Delta I_t|\le\varepsilon_{\mathrm{est}}(t)$, and bounded skew/jitter $\varepsilon_{\mathrm{sync}}$. Then for any PPT adversary,
\begin{equation}
\mathrm{Adv}^{\OPIND}_{\mathcal{A}}(\mathcal{T})
\;\le\; \sum_{t\in\mathcal{T}}\!\sqrt{2\,\Dbudget(t)}
\;+\; \sum_{t\in\mathcal{T}}\!\sqrt{2}\,\big(\varepsilon_{\mathrm{est}}(t)+\varepsilon_{\mathrm{sync}}\big).
\label{eq:opind-bound}
\end{equation}
\end{theorem}

\begin{corollary}[Uniform budgets]
\label{cor:uniform}
If $\Dbudget(t)=\Dbudget$ and $\varepsilon_{\mathrm{est}}(t)\le\bar\varepsilon_{\mathrm{est}}$ for all $t\in\mathcal{T}$ with $|\mathcal{T}|=T$, then
\[
\mathrm{Adv}^{\OPIND}_{\mathcal{A}}
\le T\sqrt{2\,\Dbudget}+T\sqrt{2}\,(\bar\varepsilon_{\mathrm{est}}+\varepsilon_{\mathrm{sync}}),
\]
i.e., advantage scales linearly in $T$ and as $\sqrt{\Dbudget}$ per interval.
\end{corollary}

\noindent\emph{Proof idea.} Pinsker bounds per-interval total variation via KL; the policy upper-bounds KL through $\Dhat_t$. A hybrid over admitted intervals gives \eqref{eq:opind-bound}; aborts simply shorten $\mathcal{T}$.

%% file: evaluation.tex
% === experiments.tex ===
\section{Experimental Evaluation}
\label{sec:evaluation}

\paragraph{Metrics.}
Unless stated otherwise, latency and power are reported \emph{per admitted interval} (conditional on non-abort); per-episode summaries appear only when explicitly labelled.

\paragraph{Thresholds (reporting \& rationale).}
Exact $(\Dbudget,\Dkill)$ used in each run are recorded in the experimentation code (versioned configs and per-run logs). We set thresholds \emph{per experiment} (e.g., by job size $n$, workload, environment) by targeting baseline quantiles of $\Dhat_t$ with a small hysteresis gap, fixing a nominal false-alarm rate and stabilising decisions under distributional shift. For Tier~II, the Tier~I thresholds are transported via a one-shot quantile-alignment map fitted on baseline data. Threshold lines on all plots match the values used at runtime (from the code/audit logs).

% --- Tier I figures (place before text to help float to top of page) ---
\begin{figure*}[!t]
  \centering
\includegraphics[width=\textwidth,height=0.24\textheight,keepaspectratio,clip,trim=0pt 0pt 28pt 0pt]{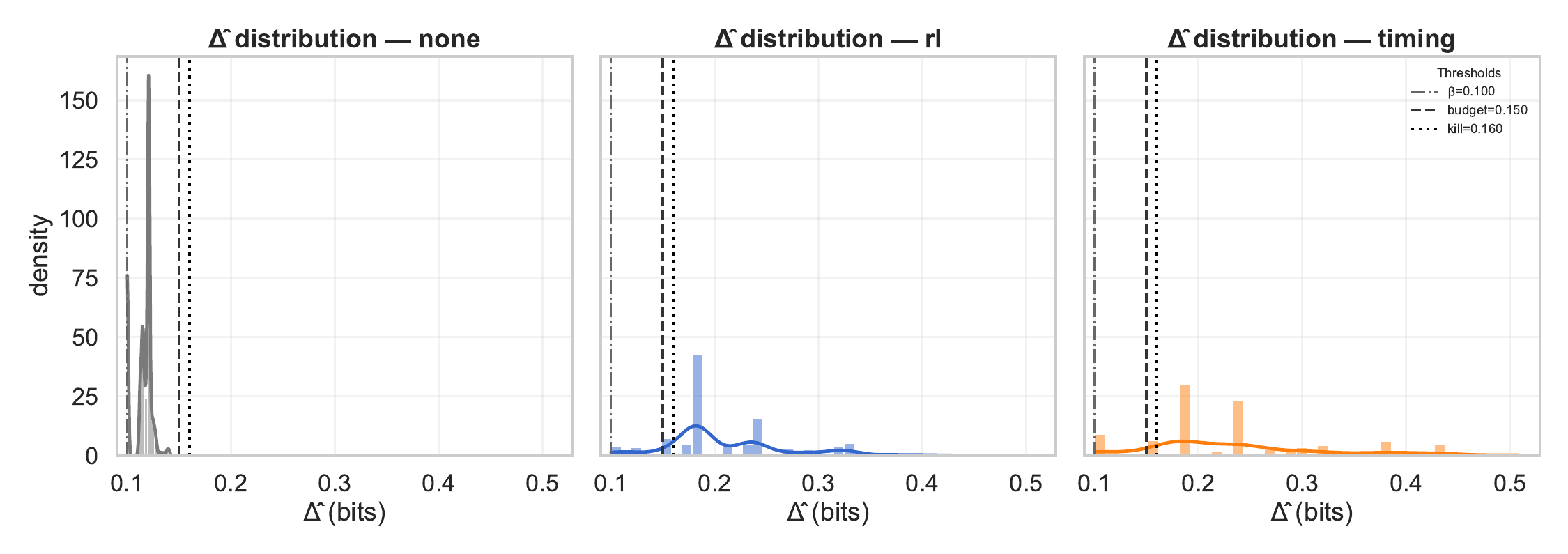}
  \caption{\textbf{Tier~I: leakage distributions.}
  Histograms of $\Dhat_t$ for baseline and adversarial workloads. Vertical lines denote the exact $(\Dbudget,\Dkill)$ used in this experiment (from the code logs). Baseline mass lies below $\Dbudget$; \texttt{rl} shifts the distribution; \texttt{timing} induces heavier right tails, approaching/exceeding $\Dkill$.}
  \label{fig:t1-hist}
\end{figure*}

% --- Tier I figure: threshold sensitivity (height-limited) ---
\begin{figure*}[!t]
  \centering
  \includegraphics[width=\textwidth,height=0.24\textheight,keepaspectratio,clip,trim=0pt 0pt 28pt 0pt]{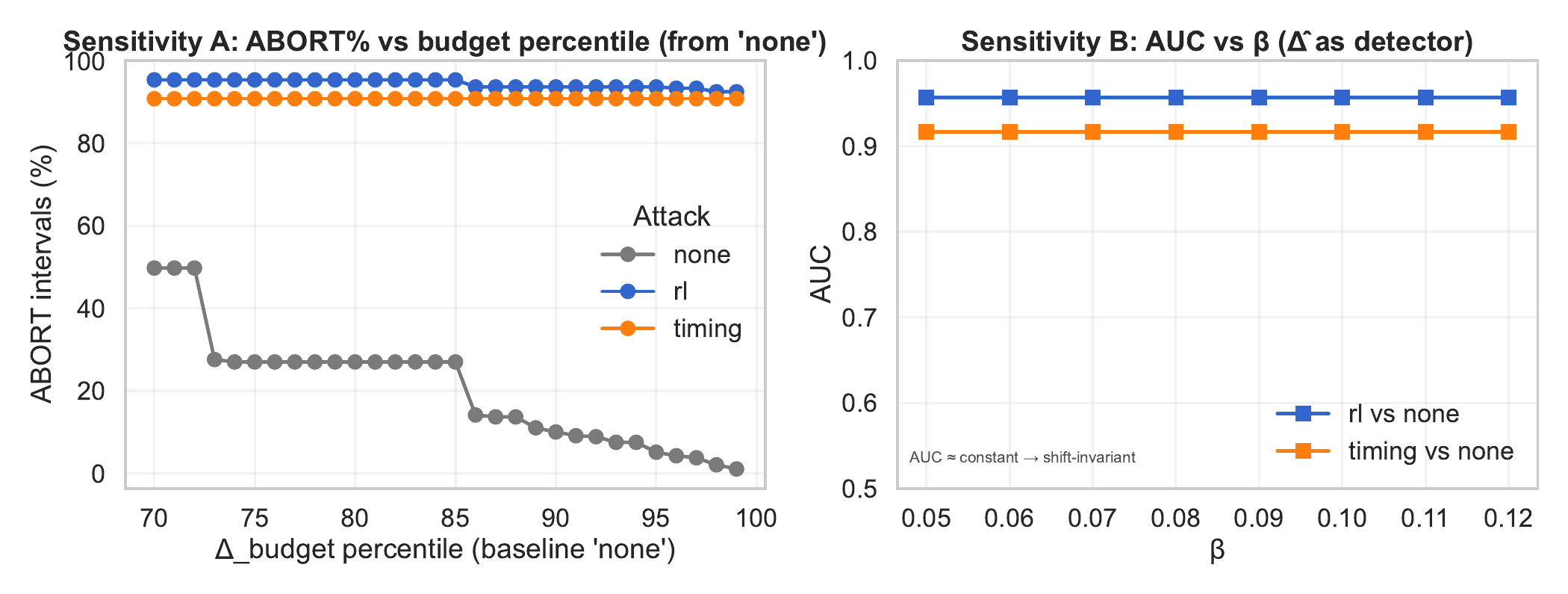}
  \caption{\textbf{Tier~I: threshold sensitivity.}
  Varying $\Dbudget$ at a fixed hysteresis gap demonstrates robust detection across operating points: tighter budgets yield higher detection at the expense of higher abort rates, as expected.}
  \label{fig:t1-sens}
\end{figure*}

\subsection{Tier~I: Adversarial Monte Carlo}
\label{ssec:tier1}

\textbf{Setup.} Job sizes $n\!\in\!\{4,8,16\}$; workloads: \texttt{none} (baseline), \texttt{rl} (batch-size modulation), \texttt{timing} (periodic burst injection). Each run uses $40$ seeds for $800$ steps, producing per-interval leakage $\Dhat_t$, policy traces, and cost metrics. Thresholds $(\Dbudget,\Dkill)$ are chosen per experiment as above. The policy mechanics (strike/cooldown with $(\Dbudget,\Dkill)$) are specified in Sec.~\ref{sec:methods}.

The leakage distributions in Fig.~\ref{fig:t1-hist} show baseline mass concentrated below $\Dbudget$, with \texttt{rl} shifting $\Dhat_t$ moderately to the right and \texttt{timing} producing the heaviest right tails that approach or exceed $\Dkill$. Vertical lines mark the exact $(\Dbudget,\Dkill)$ emitted by the code/logs for each panel, making the operating point explicit relative to the induced tails.

Sweeping $\Dbudget$ at a fixed hysteresis gap, as in Fig.~\ref{fig:t1-sens}, indicates robust detection across operating points: tightening the budget improves separability while increasing abort rates in a predictable manner, demonstrating that conclusions are not contingent on a single threshold pair.

Detailed latency and power impacts are deferred to Tier~II, where costs are reported \emph{per admitted interval}; benign overheads remain small, while adversarial pressure increases both latency and power with clear scaling in~$n$.

% --- Tier II figures (two-column) ---
\begin{figure*}[t]
  \centering
  \begin{minipage}[t]{0.32\textwidth}
    \includegraphics[width=\linewidth,clip]{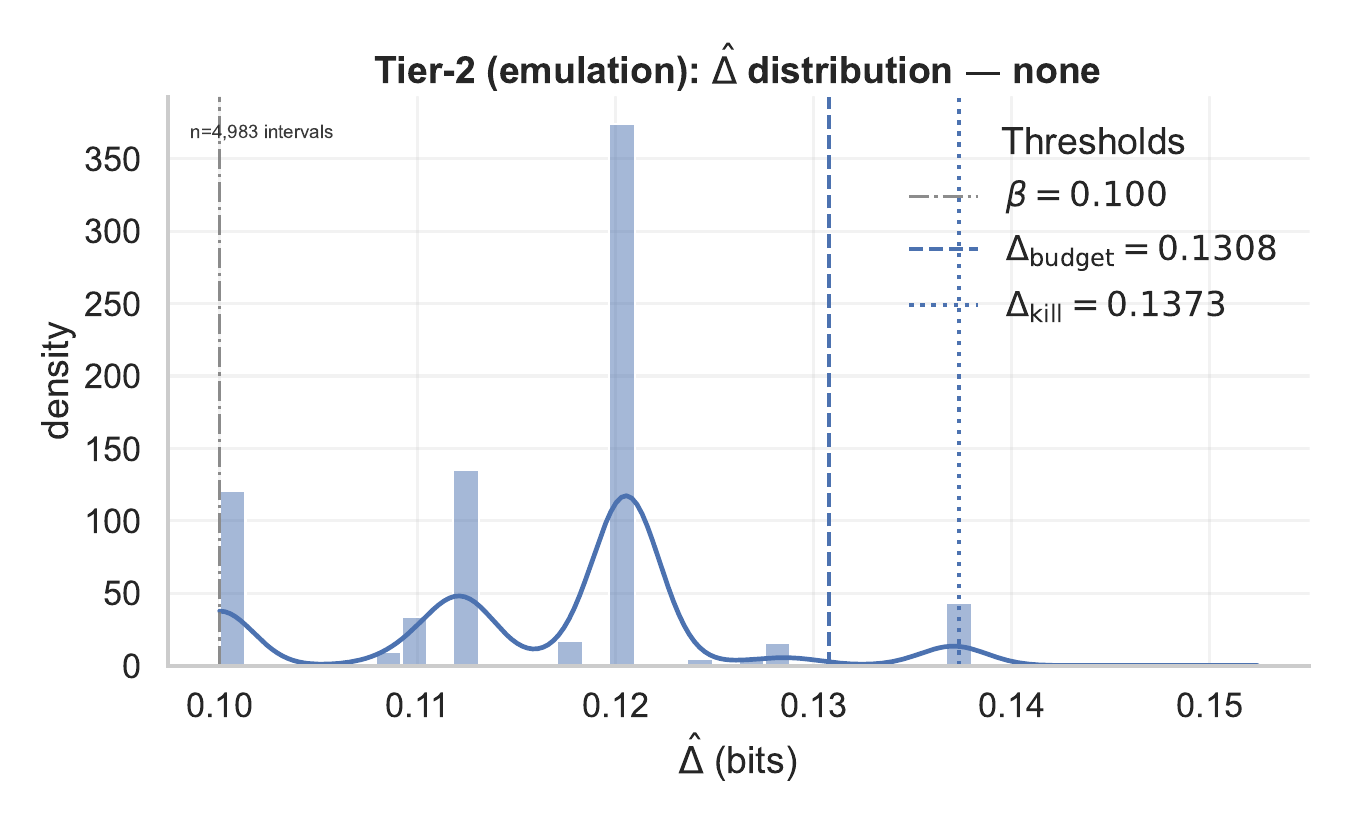}
  \end{minipage}\hfill
  \begin{minipage}[t]{0.32\textwidth}
    \includegraphics[width=\linewidth,clip]{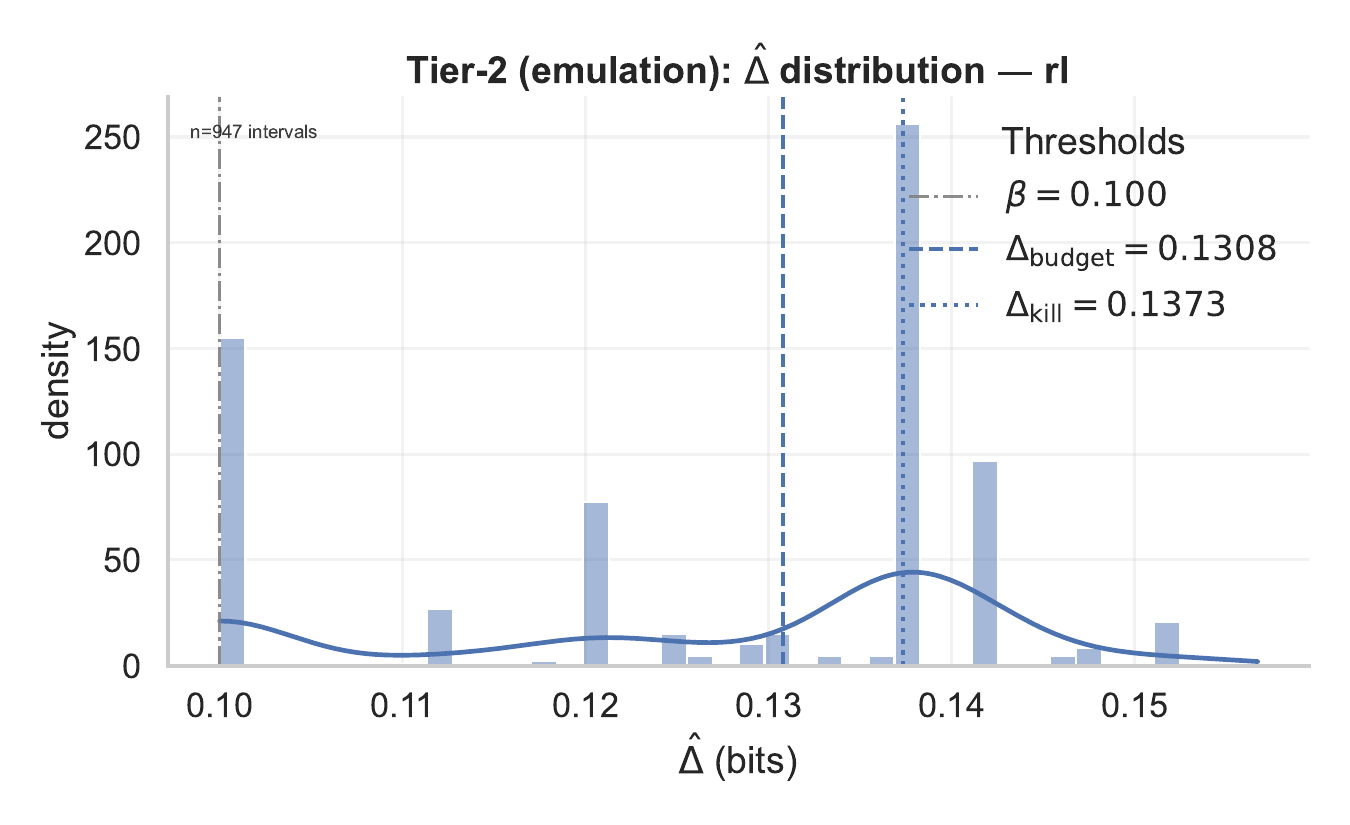}
  \end{minipage}\hfill
  \begin{minipage}[t]{0.32\textwidth}
    \includegraphics[width=\linewidth,clip]{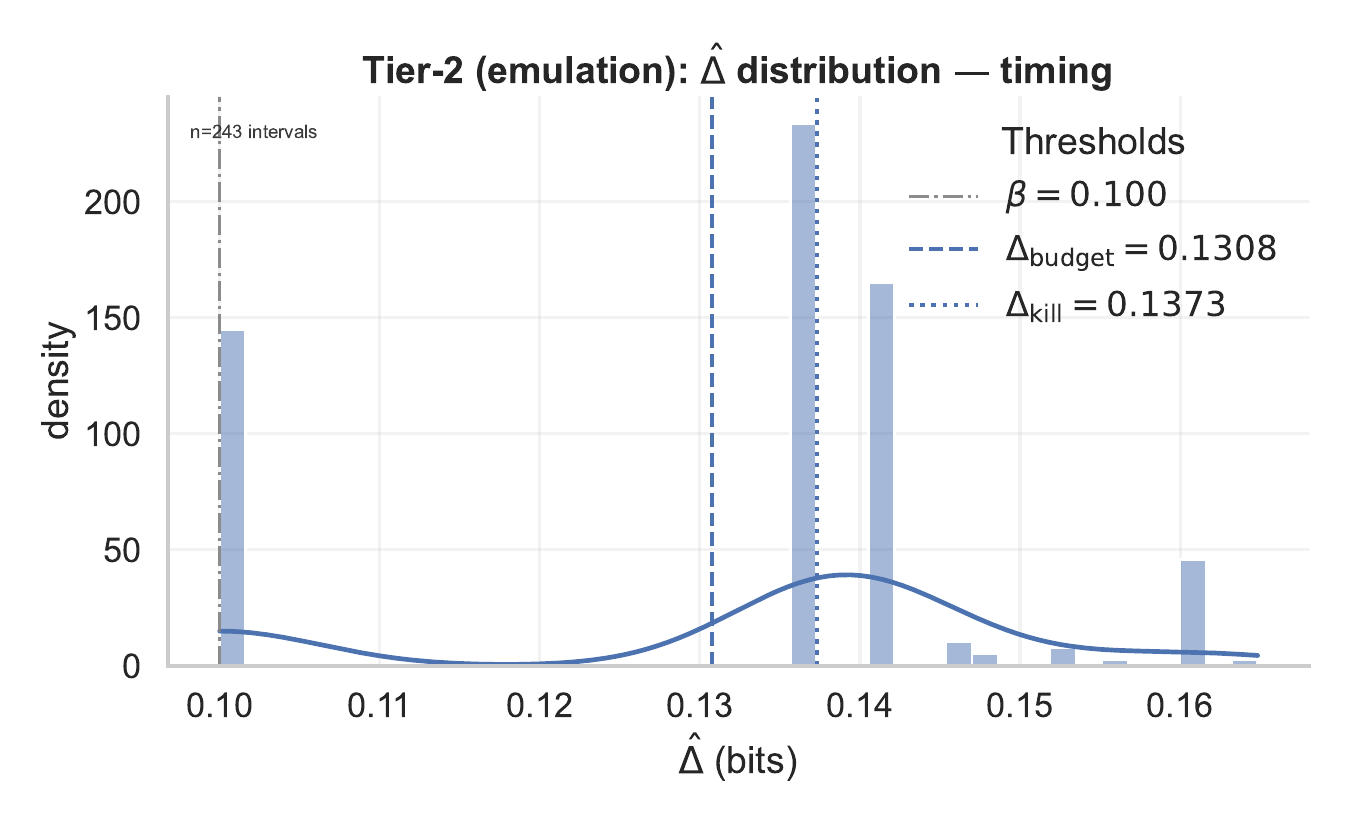}
  \end{minipage}
  \caption{\textbf{Tier~II leakage histograms (per admitted interval).}
  \emph{Left:} baseline (\texttt{none}) mass lies below $\Dbudget$. \emph{Middle:} \texttt{rl} shifts $\Dhat_t$ right. \emph{Right:} \texttt{timing} induces heavier right tails that approach/exceed $\Dkill$. Threshold lines are the exact per-experiment $(\Dbudget,\Dkill)$ emitted by the code/logs.}
  \label{fig:t2-hists}
\end{figure*}

\begin{figure*}[t]
  \centering
  \begin{minipage}[t]{0.32\textwidth}
    \includegraphics[width=\linewidth,clip]{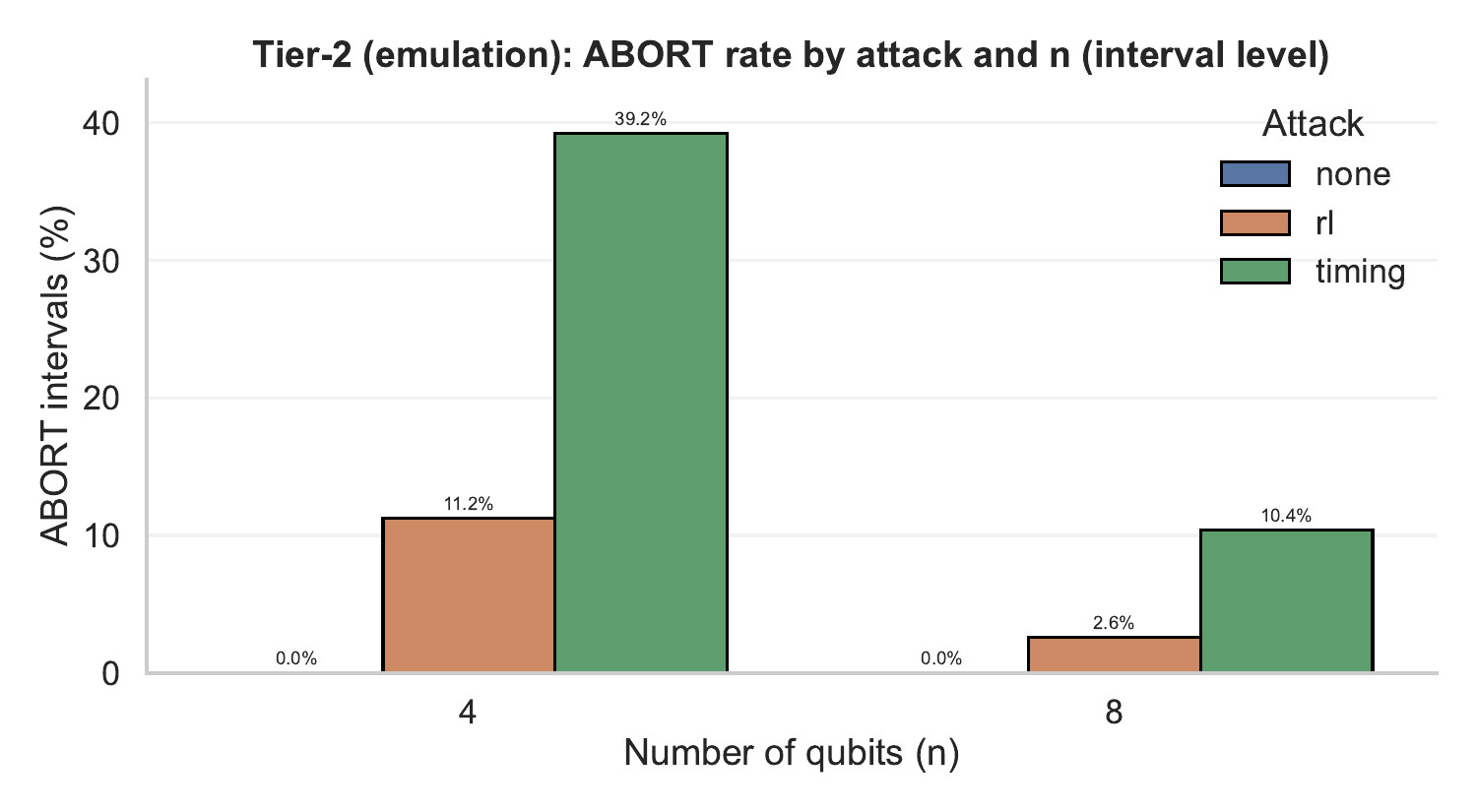}
  \end{minipage}\hfill
  \begin{minipage}[t]{0.32\textwidth}
    \includegraphics[width=\linewidth,clip]{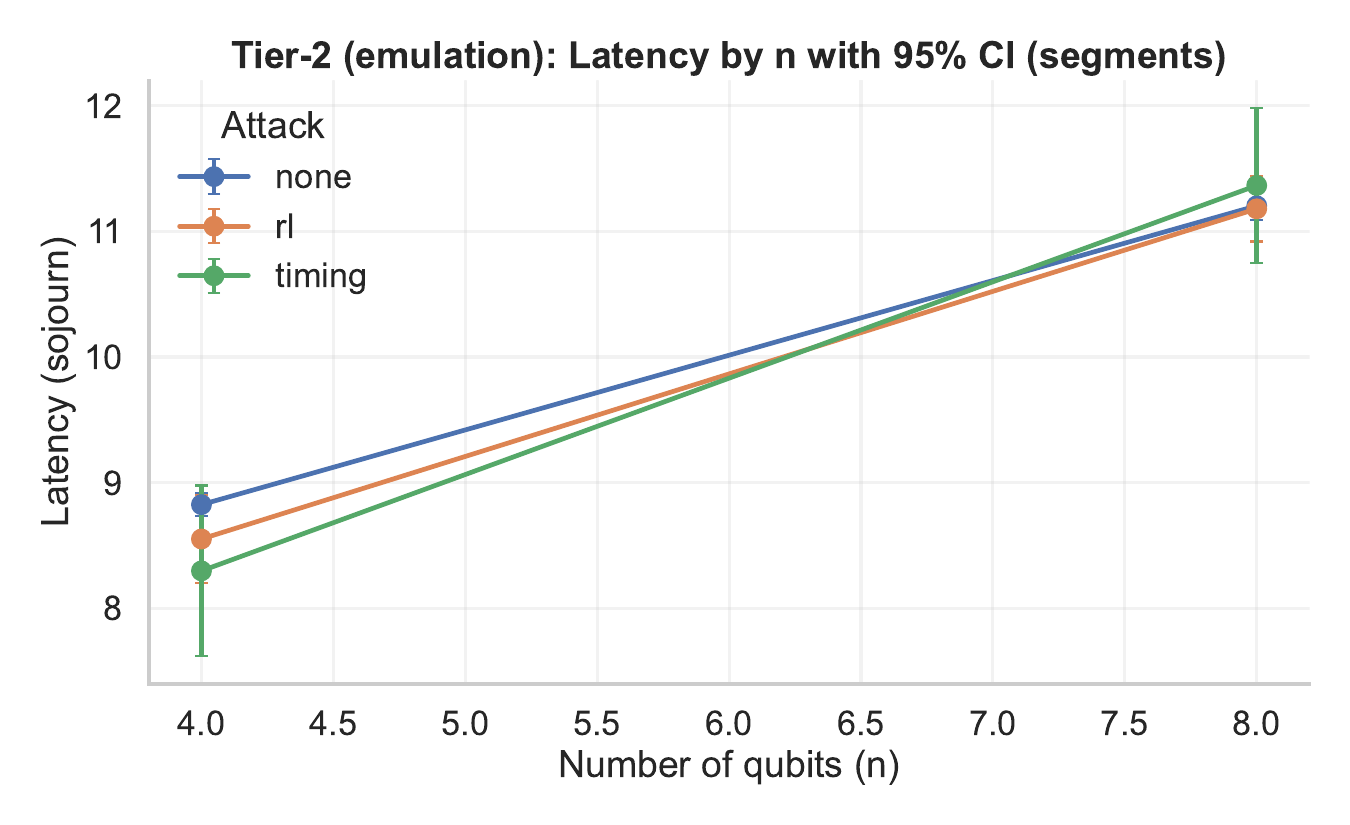}
  \end{minipage}\hfill
  \begin{minipage}[t]{0.32\textwidth}
    \includegraphics[width=\linewidth,clip]{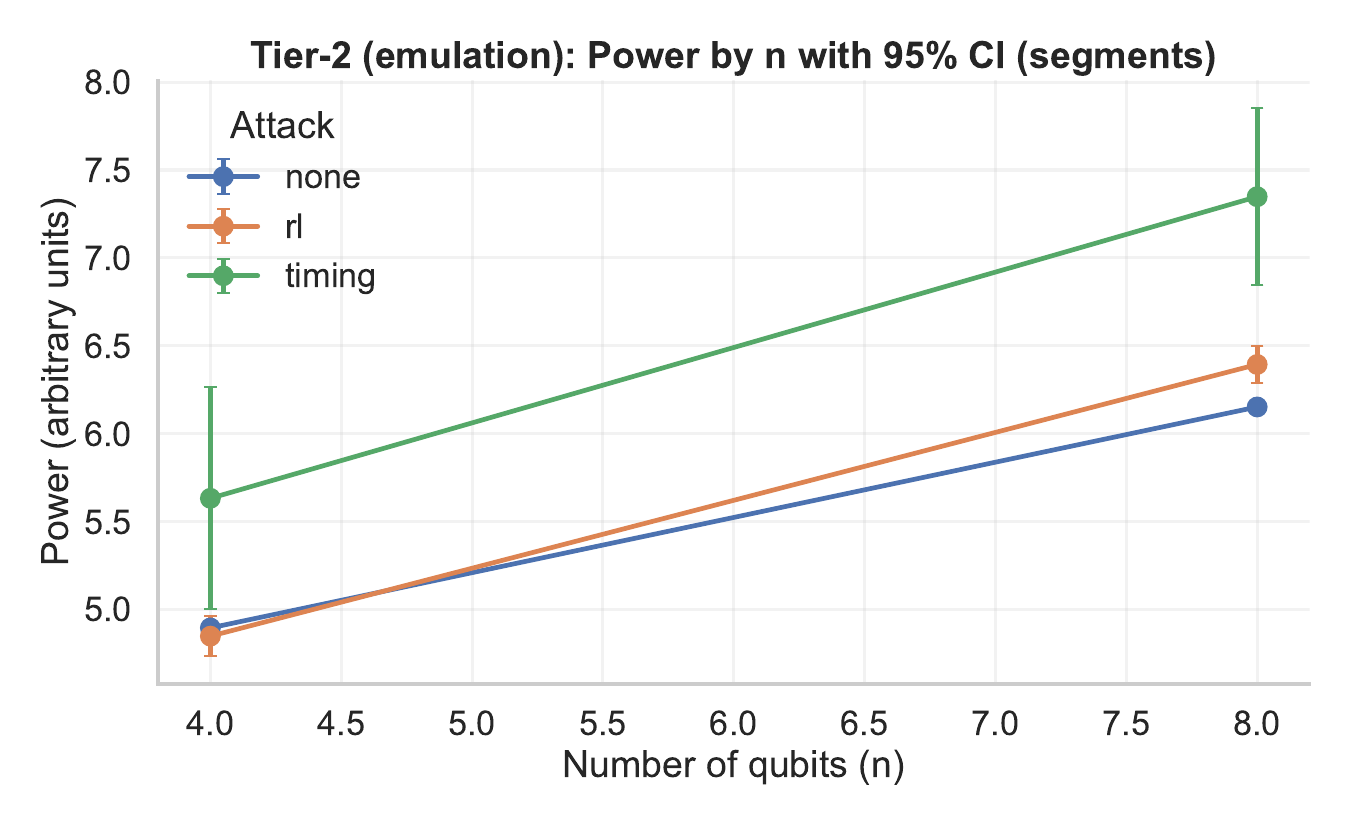}
  \end{minipage}
  \caption{\textbf{Tier~II outcomes and costs (per admitted interval).}
  \emph{Left:} ABORT rate by attack and $n\!\in\!\{4,8\}$. \emph{Middle:} latency vs.\ $n$ with 95\% CIs. \emph{Right:} power vs.\ $n$ with 95\% CIs. Adversarial workloads (especially \texttt{timing}) raise ABORT incidence and increase latency/power, consistent with the heavier leakage tails in Fig.~\ref{fig:t2-hists}.}
  \label{fig:t2-policy-costs}
\end{figure*}

\subsection{Tier~II: Cloud--Hardware Emulation}
\label{ssec:tier2}

\textbf{Setup.} We emulate realistic backend heterogeneity, FIFO queues, dispatch jitter, and CASQUE routing without vendor APIs. Workloads use $n\!\in\!\{4,8\}$ with the same seeds as Tier~I. Calibration artefacts $(\qrefdist,\beta)$ remain locked. Thresholds $(\Dbudget,\Dkill)$ are obtained per experiment via the quantile-alignment transfer from Tier~I and are recorded in the experimentation code/logs; plotted threshold lines match those artefacts.

\paragraph*{Leakage distributions.}
Per-interval histograms of $\Dhat_t$ show baseline mass concentrated below $\Dbudget$, a moderate right-shift for \texttt{rl}, and the heaviest right tails under \texttt{timing}, approaching or exceeding $\Dkill$. Each panel annotates the exact per-experiment $(\Dbudget,\Dkill)$, making the operating point clear relative to induced tails (Fig.~\ref{fig:t2-hists}).

\paragraph*{Policy outcomes and scaling.}
Interval-level ABORT rates increase under adversarial workloads and scale with $n$, with \texttt{timing} producing the highest incidence. This pattern mirrors the heavier leakage tails and indicates that the policy enforces the intended WARN/ABORT behaviour as pressure grows (left panel in Fig.~\ref{fig:t2-policy-costs}).

\paragraph*{Cost trade-offs (per admitted interval).}
Latency and power remain close to baseline under \texttt{none}, but both rise under adversarial pressure—strongest for \texttt{timing}—with growth in $n$ consistent with queue stress and routing constraints. Reporting \emph{per admitted interval} isolates performance impact from episode truncation (middle/right panels in Fig.~\ref{fig:t2-policy-costs}).

%% file: discussion_conclusion.tex
\section{Discussion and Conclusion}

\label{sec:conclusion}

NADGO delivers \emph{per-interval operational privacy} for gate-model quantum clouds by combining hardware-aware $t$-design padding, drift-robust timing randomisation, CASQUE topology-aware routing, and an online leakage estimator $\Dhat_t$ governed by locked thresholds $(\Dbudget,\Dkill)$ under the $\OPIND$ framework (Sec.~\ref{sec:security}). Across Tier~I (Monte Carlo) and Tier~II (queue/timing-aware emulation), baseline operation concentrates mass below budget with low abort incidence, while adversarial workloads shift leakage toward policy boundaries and elicit policy-compliant WARN/ABORT outcomes, with performance costs measured \emph{per admitted interval} remaining modest under benign load and rising under attack with clear scaling in $n$. The calibration artefacts are locked across tiers and thresholds are recorded per experiment for auditability. Limitations include abstraction of proprietary vendor behaviours and evaluation on small tiles; future work will address scaling, adaptive/coordinated adversaries, multi-region routing with topology-specific budgets, and succinct proofs of interval-level policy enforcement. Overall, by unifying leakage-aware compilation, drift-adaptive scheduling, and auditable enforcement, NADGO provides a practical path to certifiable operational privacy at low overhead in multi-tenant quantum clouds.